\documentclass[pra,twocolumn,floatfix]{revtex4}
\usepackage{amsmath,amssymb,amsthm}
\usepackage{graphicx}
\usepackage{subfigure}
\def\P{\ensuremath{\mathcal{P}}}
\begin{document}
\begin{abstract}
The contribution of the detector dynamics to the weak measurement is analysed. 
According to the usual theory 
[Y. Aharonov, D. Z. Albert, and L. Vaidman, Phys. Rev. Lett. {\bf 60}, 1351 (1988)] the outcome of a weak measurement with preselection and postselection can be expressed as the real part of a complex number: the weak value. By accounting for the Hamiltonian evolution of the detector, here we find that there is 
a contribution proportional to the imaginary part
of the weak value to the outcome of the weak measurement.
This is due to the coherence of the probe being essential for the
concept of complex weak value to be meaningful. 
As a particular
example, we consider the measurement of a spin component and find
that the contribution of the imaginary part of the weak value is
sizeable. 
\end{abstract}
\title{Weak measurement: Effect of the detector dynamics}
\author{Antonio \surname{Di Lorenzo}}
\author{J. Carlos Egues}
\affiliation{Departamento de F\'{\i}sica e Inform\'{a}tica, 
Instituto de F\'{\i}sica de S\~{a}o Carlos, \\
Universidade de S\~{a}o Paulo, 13560-970 S\~{a}o Carlos, S\~{a}o
Paulo, Brazil}
\date{\today}

\maketitle
\section{Introduction}
The concept of weak value was introduced in \cite{AAV}. It is the
complex number $A_{w}=\langle S_{f}|\Hat{A}|S_{i}\rangle/\langle
S_{f}|S_{i}\rangle$ in terms of which one can express $\langle
A\rangle$, the average result of a measurement of an observable
$\Hat{A}$ preceded by a preparation in the state $|S_{i}\rangle$ and
followed by a postselection of the state $|S_{f}\rangle$, provided
that the interaction between the system and the detector, which we
shall call the probe as a reminder of its quantum nature, is weak
enough compared to the coherence scale of the latter \cite{Duck}.

Under the assumptions of \cite{AAV}, for a weak analogue of an ideal
von Neumann measurement, the average value is given by $\langle A\rangle={Re}(A_{w})$. 
A surprising result is that this average
value can lie well outside the range of the eigenvalues of
$\Hat{A}$. This fact has been confirmed experimentally \cite{Ritchie,Parks,Pryde} in optics. 
Also, the formalism of the weak value was proved to describe some relevant phenomena 
in telecom fibers \cite{Brunner2003}, and to be connected with the response function 
of a system \cite{Solli2004}. 
The possibility of performing a weak measurement in solid state systems is 
currently being investigated \cite{Romito}.
In Ref. \cite{AAV} the initial state of the probe is assumed a pure
gaussian state, with a special choice of the phase, and the free
evolution of the probe is neglected. Since the coherence of the
probe is an essential requisite for the weak value to be
significant, and since the Hamiltonian evolution induces a relative
phase between different components of the state of the probe, the
latter assumption seems unrealistic, especially for a measurement
lasting a finite time.

In this paper we calculate $\langle A\rangle$ for any initial state
of the probe and for any interaction strength 
[Eqs. (\ref{eq:instgenprob}),(\ref{eq:genweakval})]. In the limit of a
weak interaction, we show that including the free evolution of the
probe gives rise to a contribution  $\propto {Im}(A_{w})$ to
$\langle A\rangle$ [Eq.~(\ref{eq:genweakval2})];
this, generally, does not change the main property of the weak
measurement, namely that $\langle A\rangle$ can lie well outside the
spectrum of $\Hat{A}$. We then consider, as a special example, a
probe prepared in a general gaussian state, including the state
assumed in \cite{AAV} as a particular case, and we provide
additionally an expression for the variance $\langle \Delta
A^2\rangle$, Eq.~\eqref{eq:spread}. Finally, we take $\Hat{A}$ to be
a spin component, as a simple illustration; we discuss the regime
where the weak value does not apply, providing formulas for the
extrema of $\langle A\rangle$, $\langle \Delta A^2\rangle$ as a
function of the postselection
[Eqs.(\ref{eq:maxpos},\ref{eq:maxval},\ref{eq:spreadmin},\ref{eq:spreadmax})].

\section{Measurement statistics with Pre and Postselected states}
Let us consider a quantum system prepared, at time $t_{i}$, in a pure
state $|S_{i}\rangle$ (preselection). We denote the Hamiltonian of the
system by $H_{sys}$.
The system interacts, at time $T_{i}\ge t_{i}$, with another quantum
system, the probe, through
$H_{int} = -g(t) \lambda \Hat{q} \Hat{A}$, where $\Hat{A}$ is an
operator on the system's Hilbert space, $\Hat{q}$ on the probe's,
and $g(t)$ is a function vanishing outside a finite interval
$[T_{i},T_{f}]$, with $\int g(t) dt = 1$.
For the measurement to be ideal, if $\Hat{A}$ is not conserved, the
interaction must be instantaneous, $T_{f}=T_{i}$, $g(t)=\delta(t-T_{i})$;
otherwise, if $\Hat{A}$ is conserved, i.e.
$\left[\Hat{A},\Hat{H}_{sys}\right]=0$,  the interaction can last a
finite time, and the measurement is a non-demolition one \cite{Khalili}.
The probe is prepared, at time $t_{p}\le T_{i}$, in a state described
by the density matrix $\Hat{\rho}$, and its free evolution is
governed by $H_{p}(\Hat{p})$, where $\Hat{p}$ is the conjugate
observable of $\Hat{q}$. The operator $\Hat{p}$ is the observable of
the probe that carries information about the measured quantity
$\Hat{A}$. We notice that, in order for the measurement to be ideal,
$\Hat{p}$ must be conserved during the free evolution of the probe,
and change only due to the interaction with the observed system.
\begin{figure}[hb]
\includegraphics[height=1in,width=3in]{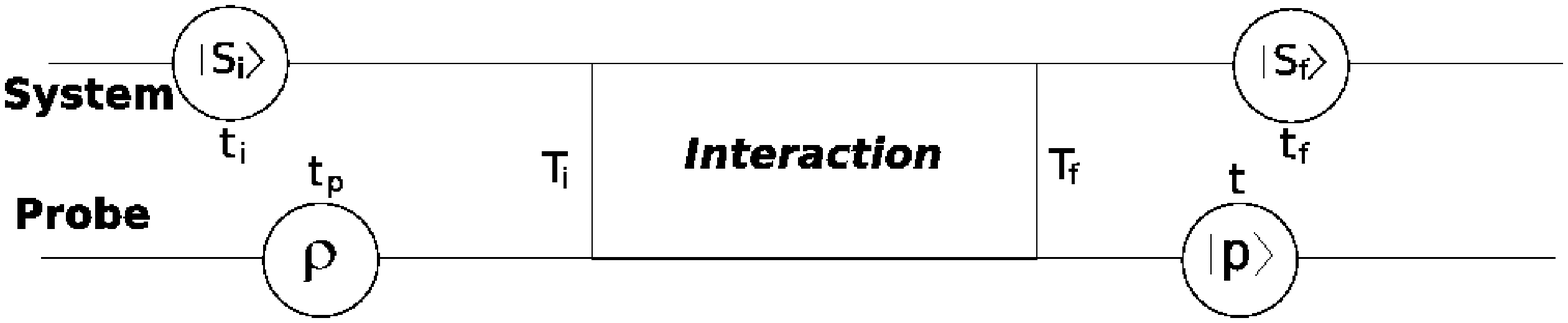}
\caption{\label{fig:scheme} A schematic view of the measurement with
pre- and post-selection, the horizontal direction representing
increasing time. }
\end{figure}
At time $t_{f}\ge T_{f}$, a sharp measurement \footnote{By sharp
measurement, we mean a measurement satisfying the first half of
Born's rule, i.e. the outcome of which corresponds to an eigenvalue
of the observable under detection. This last measurement need not be
a projective one, since it is immaterial what the state of the
system is afterwards.} 
of an observable $\Hat{S}_{f}$ of the system is made, giving an outcome $S$, corresponding to the
eigenstate $|S\rangle$. At time $t\ge T_{f}$ a sharp measurement of
the observable $\Hat{p}$ is made on the probe. Since $\Hat{p}$ is
conserved during the free evolution of the probe, this value will
not depend on the time $t$. The observed value of non-conserved
quantities, by contrast, would depend on $t$ \footnote{In classical
mechanics one could deterministically predict the value that an
observable $O$ will have at time $t$ in the absence of interaction
with the measured system, and hence infer something about the
measured system from the value of $O$ observed in the presence of
interaction. Due to the stochastic nature of Quantum Mechanics, this
is no longer possible: the uncertainty on a non-conserved quantity
will generally spread with time. For this reason, in an ideal
measurement, the pointer variable is required to be conserved.}.
Finally, only those trials in which the last measurement on the
system gave an arbitrarily fixed outcome $S=S_{f}$ will be selected
(postselection). The procedure detailed above describes a
measurement with pre- and post-selection. In Fig.\eqref{fig:scheme}
we provide a sketch of the procedure.

The joint probability of observing the outcome $p$ for the probe, at
any time $t\ge T_{f}$ and $S_{f}$ for the system, at time $t_{f}$, is
given by Born's rule
\begin{align}
\nonumber
&\P(p,S_{f}|\rho, S_{i}) = \int dp_0 dp_0' e^{i\left[H_{p}(p_0')-H_{p}(p_0)\right] (T_{i}-t_{p})/\hbar}\\
&\times \rho(p_0,p_0') \langle \tilde{S}_{f}|\langle p| \mathcal{U}_{T_{i},T_{f}}
|\tilde{S}_{i}\rangle |p_0\rangle  \langle \tilde{S}_{i}| \langle p_0'|
\mathcal{U}_{T_{i},T_{f}}^\dagger |\tilde{S}_{f}\rangle |p\rangle
\end{align}
where $\mathcal{U}$ is the time evolution operator for
$H_{sys}+H_{p}+H_{int}$, we introduced $\rho(p,p')$ the probe density matrix in the $|p\rangle$ basis,
and
$
|\tilde{S}_{i,f}\rangle:=\exp{\left\{-i\Hat{H}_{sys}(T_{i,f}-t_{i,f})/\hbar\right\}}|S_{i,f}\rangle
$. 
After introducing twice the identity resolved in terms of the
eigenstates of $\Hat{A}$, we obtain
\begin{align}
\nonumber \P(p,S_{f}|\rho, S_{i}) =& \sum_{a,a'} \rho(p-\lambda
a,p-\lambda a') e^{-i\left[\Phi_{a}-\Phi_{a'} \right]}
\\
\label{eq:instgenprob} 
&\times 
\langle \tilde{S}_{f}| a \rangle \langle a|\tilde{S}_{i}\rangle \langle
\tilde{S}_{i}| a'\rangle \langle a'|\tilde{S}_{f}\rangle ,
\end{align}
with $a, |a\rangle$ the eigenvalues and eigenvectors of $\Hat{A}$
and
\begin{equation}\label{eq:genhamphase}
\Phi_a:=\frac{1}{\hbar}\int_{t_{p}}^{T_{f}}\!\!\!ds\
H_{p}\!\left(p-\lambda\; a\!\int_s^{T_{f}} ds' g(s')\right).
\end{equation}
For an instantaneous interaction, $T_{f}$ in Eq.~\eqref{eq:genhamphase} and elsewhere in this paper
should be interpreted as being a time infinitesimally later than the interaction. 
In deriving Eq.~\eqref{eq:instgenprob}, we exploited
\begin{align}
\nonumber 
\langle p|\mathcal{T}\!e^{\!-i\!\int_0^t\!ds\!
\left[H_{p}\!(\Hat{p})\!-\!f\!(s)\Hat{q}\right]\!/\!\hbar}  |p_0\rangle
&=\delta\!\left(p\!-\!p_0\!-\!\int_0^t ds f(s)\right)
\\
&\times\!e^{-i\!\int_0^t\!ds\!H_{p}\left(\!p\!-\!\int_s^{t}\!ds' f\!(s')\!\right)\!/\!\hbar}.
\end{align}
We notice that if no postselection were made (i.e. if one would sum over the final states $|S_{f}\rangle$), 
the off-diagonal elements of $\rho(p,p')$ would not contribute to 
Eq. \eqref{eq:instgenprob}. 

The conditional probability of obtaining outcome $p$, given that the
state has been postselected in $S_{f}$, is
\begin{align}
\label{eq:genweakprob} 
\P(p|S_{f},\rho,S_{i}) =
\frac{\P(p,S_{f}|\rho, S_{i})}{\int dp' \P(p',S_{f}|\rho, S_{i})},
\end{align}
where we applied Bayes' rule, and the expected value inferred for
the system through observation of the probe
\begin{equation}
\label{eq:genweakval}
 \langle A\rangle := \frac{\langle p \rangle}{\lambda}=
\frac{\int\!d\!p\, (p/\lambda) \P(p,S_{f}|\rho,S_{i})}
{\int\!d\!p'\P(p',S_{f}|\rho, S_{i})}.
\end{equation}

Since what matters is the deviation of the pointer $p$ from its
unperturbed expected value $\int dp \ p \rho(p,p)$, we can set the
latter to be zero without loss of generality. We notice that the
inference assigning the quantity $A=p/\lambda$ to the system when
observing the probe to have the value $p$ is valid only when
initially the probe has a precise enough value of $p$ close to zero.
We can, however, keep assigning the value to the system even when
the probe is not sharply prepared around $p=0$, and say that we
observed a value $A$ for the measured system (and correspondingly we
shall define a probability $\P(A):=\lambda \P(p/\lambda)$). This
value is not necessarily one of the eigenvalues of $\Hat{A}$, and,
as shown in \cite{AAV}, it can even lie outside the range
$[a_{min},a_{max}]$ (for this reason we are indicating with $a$ the
eigenvalues of $\Hat{A}$ and with $A$ the outcome of each individual
measurement).

\section{Weak Measurement}
A measurement is weak when the coupling $\lambda$ is small compared
to the coherence length of the probe, i.e. to the range of $|p-p'|$
within which $\rho(p,p')$ vanishes. This can be evinced from Eq. \eqref{eq:instgenprob}.

In the following, we shall assume that $\rho(p,p')$ is analytic in a
neighborhood of $p=p'$. Then, to lowest order in $\lambda$, the
denominator in Eq.\eqref{eq:genweakval} is $ \int dp
\P(p,S_{f}|\rho,S_{i}) =|\langle \tilde{S}_{f} |\tilde{S}_{i}\rangle|^2 $.
Before analysing the numerator in Eq.\eqref{eq:genweakval}, we
rewrite $\rho(p,p') = F(p,p') \exp{\{i\alpha(p,p')\}}$, with
$F,\alpha$ symmetric and antisymmetric real functions, respectively.
We have then that the numerator in Eq.\eqref{eq:genweakval} is
\begin{align}
\nonumber &\int dp \frac{p}{\lambda} \P(p,S_{f}|\rho,S_{i})\simeq\!-\!
\sum_{a,a'} \langle \tilde{S}_{f}| a \rangle \langle
a|\tilde{S}_{i}\rangle \langle \tilde{S}_{i}| a'\rangle \langle
a'|\tilde{S}_{f}\rangle
\\
\nonumber & \int dp \ \frac{p}{2} \left\{
(a+a') \frac{d\P(p)}{dp}
-i (a-a')  \P(p) G(p)
\right\}\\
\nonumber =& {Re}\!\left(\langle \tilde{S}_{f}| \Hat{A}
|\tilde{S}_{i}\rangle \langle \tilde{S}_{i}|\tilde{S}_{f}\rangle \right)
\!-\!\overline{p G(p)}{Im}\!\left(\langle \tilde{S}_{f}|
\Hat{A} |\tilde{S}_{i}\rangle \langle
\tilde{S}_{i}|\tilde{S}_{f}\rangle\right) ,
\end{align}
where $\P(p):=F(p,p)$ is the initial distribution of the $p$
observable of the probe,
\begin{align}
\nonumber &G(p)=\frac{2\Delta t}{\hbar} \frac{d
H_{p}(p)}{d p}-2\left.\frac{\partial \alpha(p,p')}{\partial
p}\right|_{p'=p},
\\
\label{eq:time} &\Delta t=\int_{t_{p}}^{T_{f}} ds \int_s^{T_{f}} ds'
g(s')\ ,
\end{align}
and the bar symbol denotes the average over $\P(p)$. After
introducing the weak value, i.e. the complex number $A_{w}:={\langle
\tilde{S}_{f}| \Hat{A} |\tilde{S}_{i}\rangle} /{\langle
\tilde{S}_{f}|\tilde{S}_{i}\rangle}$, the average value is
\begin{equation}\label{eq:genweakval2}
\langle A\rangle  \simeq  \langle A\rangle_0  :={Re}\!\left\{
A_{w} \right\} -\overline{p G(p)}\ {Im}\!\left\{ A_{w} \right\}.
\end{equation}
Eq.\eqref{eq:genweakval2} holds as far as the product of the
prepared and the postselected state is larger than the first
nonvanishing contribution in the $\lambda$ expansion for the
denominator. In the latter case, one should keep the latter
contribution as well.

\section{Comparison with previous results}
We notice that the contribution of the imaginary part has been
generally overlooked in the literature, due to the neglecting of the
Hamiltonian of the probe and to the choice of a very special phase
$\alpha(p,p')=0$. On the other hand, it has been proved \cite{AAV},\cite{AV} that
observing the $\Hat{q}$ variable of the probe one gets an average
value which is proportional to the imaginary part of $A_{w}$. This is
true only if one neglects the time evolution of the probe from
preparation to observation. When this evolution is accounted for,
the observed value of $\Hat{q}$ depends also on the details of the
free Hamiltonian of the probe and on the time of observation. To the
best of our knowledge, the first paper to point out that 
${Im} (A_{w})$ contributes to $\langle A \rangle$ was reference
\cite{Jozsa}. There, however, the readout variable $\Hat{p}$ (which
in the notation of \cite{Jozsa} is actually $\Hat{q}$) is not
conserved during the free evolution of the probe. Thus the results
presented in \cite{Jozsa} hold only if the system-probe interaction
is instantaneous, and if the probe is read immediately after the interaction.
Indeed, if the interaction is instantaneous, the probe is prepared at time $t_{p}$ and 
it is observed at time $t$, the central result, Eq.(17), of Ref.~\cite{Jozsa} should be substituted by 
(we use our notation $\Hat{p}\leftrightarrow \Hat{q}$)
\begin{align}\nonumber
&\langle M\rangle \simeq 
\langle 
\Hat{M}(t)
\rangle_{p} +
i\lambda \biggl[{Re}A_{w} 
\langle
\left[\Hat{M}(t),\Hat{q}(T_{i})\right]
\rangle_{p}
\\
&-{Im}A_{w} \left( 
\langle
\left\{\Hat{M}(t),\Hat{q}(T_{i})\right\}
\rangle_{p}
-2
\langle 
\Hat{M}(t)
\rangle_{p}
\langle
\Hat{q}(T_{i})
\rangle_{p}
\right)\biggr],
\end{align}
where $[\cdot,\cdot]$ ($\{\cdot,\cdot\}$) denotes (anti)commutator, 
$\langle \Hat{M} \rangle_{p} = {Tr}\{\Hat{\rho} \Hat{M}\}$ is the trace taken 
with the probe density matrix at time $t_{p}$, 
and $\Hat{M}(t)$ is the probe observable $\Hat{M}$ evolved with $H_{p}$ in the interval $[t_{p},t]$. 
Generally, with a probe Hamiltonian $H_{p}=\Hat{p}^2/2M_{p}+V(\Hat{q})$, one can no longer link the contribution of ${Im}\{A_{w}\}$ to $\langle p\rangle$ 
with the derivative of the variance of $q$, unless $t=T_{i}$. 
However, if the probe Hamiltonian is $\Hat{H_{p}}=\Hat{p}^2/2M_{p}$, 
we have that $\Hat{p}(t)=\Hat{p}(T_{i})$ and thus  
\begin{equation}
\langle p\rangle\simeq \lambda {Re}\{A_{w}\}
-\lambda \frac{M_{p}}{\hbar} \beta {Im}\{A_{w}\}, 
\end{equation}
where $\beta := \left. {d{Var}q(t)}/{dt}\right|_{t=T_{i}}$.
The formula agrees with the more general Eq.~\eqref{eq:genweakval2}, since for a generic $H_{p}(p)$ 
\[\frac{d{Var}q(t)}{dt} 
=\hbar \overline{\frac{dH_{p}(p)}{dp}G(p)}-\hbar \overline{\frac{dH_{p}(p)}{dp}}\overline{G(p)},\]
which reduces to 
${d{Var}q(t)}/{dt}=\hbar\overline{pG(p)}/M_{p}$.
For a generic Hamiltonian $H_p(\Hat{p})$, 
if instead of observing $\Hat{p}$, one observes the ``velocity'' operator 
$\Hat{V}=V(\Hat{p}):=dH_{p}(\Hat{p})/dp$, one has 
\begin{align}\label{eq:genjozsa}
&\langle V\rangle \!\simeq\! \langle \Hat{V}\rangle_{p} \!+\!
\lambda \langle
\frac{d \Hat{V}}{d\Hat{p}}
\rangle_{p}
{Re}\{A_{w}\} 
\!-\!\lambda \beta {Im}\{A_{w}\} 
.
\end{align}

\section{Probe prepared in a mixed gaussian state}
So far, in the literature on the weak measurement, the probe was
assumed to be prepared in a pure gaussian state. The corresponding
density matrix $\rho(p,p')$ is characterized by the identity between
the scale in $|p-p'|$ over which its off-diagonal elements vanish
(the coherence length scale) and the scale over which the diagonal
elements decay going away from the zero value (the classical
uncertainty spread in $p$). We shall consider a more general
gaussian distribution
\begin{equation}\label{eq:gengauss}
\rho(p,p')\!=\! \frac{ e^{- \left\{\!(p+p')^2/8\Delta P^2
\!+(p-p')^2/8\delta p^2 -i\left(p-p'\right)/2 p_\phi\!\right\}
}}{\sqrt{2\pi}\Delta P}\!.
\end{equation}
Here $\Delta P$ is the initial spread and $\delta p$ the coherence
scale. Positive semidefineteness requires $\delta p \le \Delta P$.
We assumed a phase linear in $p$, with $p_\phi$ a scale.
The linear phase such chosen defines the center of the Wigner
function in the coordinate $Q$, $Q_0 :=\hbar/2 p_\phi$. We take a
quadratic Hamiltonian for the free probe $H_{p}(\Hat{p}) =
\Hat{p}^2/2M_{p}$, and we define a further scale $p_H:=\sqrt{\hbar
M_{p}/2\Delta t}$, with $\Delta t$ defined by Eq.\eqref{eq:time}. We
stress that the presence of $\sqrt{\hbar}$ can make this scale the
smallest one. We have then that the average detected value of the
observable $\Hat{A}$ is given by
Eq.~\eqref{eq:genweakval2} with $\overline{p G(p)} =\Delta
P^2/p_H^2:=\kappa^2$. The results of Ref.~\cite{AAV} are recovered
for $\Delta P=\delta p$, and $p_H,p_\phi\to \infty$.
We also provide an expression for the variance of $A$
\begin{align}
\label{eq:spread} 
\langle\Delta A^2 \rangle \simeq \frac{\Delta
P^2}{\lambda^2} &+ \frac{1}{2}\left(1- \kappa^4\right)
{Re}\Delta A^2_{w}
- \kappa^2
{Im}\Delta A^2_{w}
,
\end{align}
where we introduced $\Delta A^2_{w}:=(A^2)_{w}-(A_{w})^2$, with $(A^2)_{w} :=\langle S_{f}|\Hat{A}^2|S_{i}\rangle/\langle S_{f}|S_{i}\rangle$. 
We notice that there is always a large contribution $\Delta
P^2/\lambda^2$, due to the initial spread in $p$.
The calculated variance differs from Eqs.~(24,25) of Ref. \cite{AV}: 
even for $\kappa=0$ (which is the limit considered in \cite{AV}), 
Eq.~\eqref{eq:spread} allows $\langle \Delta A^2\rangle < \Delta P^2/\lambda^2$.
We also stress that if the last two terms in Eq.\eqref{eq:spread}
take a small value, this does not imply that a single measurement
can reveal the weak value: $A$ is inferred from the observed $p$ of
the probe through $A=p/\lambda$; since $p$ has a spread of order
$\Delta P$, the value of $A$ observed in each individual measurement
will vary with a spread of order $\Delta P/\lambda$, which is large
by hypothesis.

\section{Illustration: weak measurement of a spin component}
As a specific example, we consider a measurement of spin components.
We assume that the interaction between the spin and the probe lasts
a finite time $T$, that $g(t)=1/T$ during the interaction and zero
otherwise, and that the probe is prepared in the state $\rho$ of
Eq.~\eqref{eq:gengauss} immediately before the beginning of the
interaction. Then Eq.~\eqref{eq:time} gives $\Delta t=T/2$.
We take the spin to have been preselected in the state up along
direction $\mathbf{n}_{i}$ and postselected in the state up along a
direction $\mathbf{n}_{f}$, while
$\Hat{A}=\mathbf{n}\cdot\Hat{\boldsymbol{\sigma}}$.
Then we obtain from
Eqs.(\ref{eq:genweakprob},\ref{eq:instgenprob})
\begin{align}
\nonumber &
\P(p|S_{f},\rho,S_{i})
= \frac{1}{\sqrt{2\pi}\Delta P N}
\\
& \nonumber \frac{1}{2} \sum\limits_{\sigma=\pm 1} \Biggl\{
\left(1+\sigma\mathbf{n}\cdot\mathbf{n}_{i}\right)
\left(1+\sigma\mathbf{n}\cdot\mathbf{n}_{f}\right)
e^{-(p-\sigma\lambda)^2/2\Delta P^2}
\\
\nonumber &+\biggl[ (\mathbf{n}\!\times\!\mathbf{n}_{i})\cdot
(\mathbf{n}\!\times\!\mathbf{n}_{f}) \cos{\left[\lambda G(p)\right]}
\\
&-\mathbf{n}\cdot(\mathbf{n}_{i}\!\times\!\mathbf{n}_{f})
 \sin{\left[\lambda G(p)\right]}
\biggr]
 e^{-p^2/2\Delta P^2-\lambda^2/2\delta p^2}
\Biggr\},
\\
\nonumber &N=1+\mathbf{n}_{i}\cdot\mathbf{n}_{f}
+e^{-\lambda^2/2\nu^2}\sin{\left(\frac{\lambda}{p_\phi}\right)}
\mathbf{n} \cdot(\mathbf{n}_{i}\times\mathbf{n}_{f})
\\
&\label{eq:normalizing} -\left[1-e^{-\lambda^2/2\nu^2}
\cos{\left(\frac{\lambda}{p_\phi}\right)}\right]
(\mathbf{n}\times\mathbf{n}_{i})\cdot (\mathbf{n}\times\mathbf{n}_{f}) ,
\end{align}
with $\nu:=[1/\delta p^2+\kappa^4/\Delta P^2]^{-1/2}\le \Delta
P/\sqrt{1+\kappa^4}$.
\begin{figure}[t!]
\includegraphics[width=3in,height=2in]{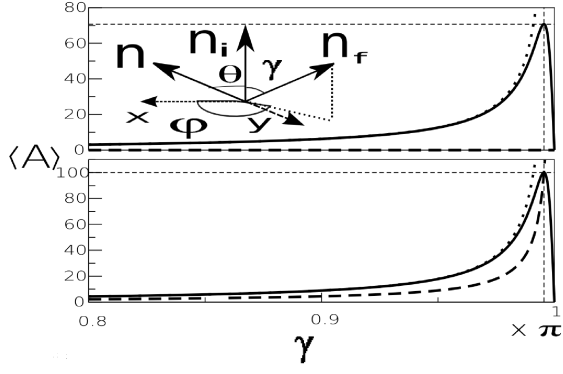}
\caption{\label{fig:example} 
$\langle A\rangle$ as a function of
$\gamma$ for $\theta=\pi/2$, and $\phi=\pi/2$ (top figure),
$\phi=\phi^*=\pi/4$ (bottom figure).
(The inset shows how the angles were
defined.) The full line depicts the exact
value, the dotted one depicts the approximate value including the
contribution of ${Im}\{A_{w}\}$, and the dashed line the
contribution of ${Re}\{A_{w}\}$. The thin vertical and
horizontal lines correspond to $\gamma^*=\pi-\lambda/\nu$, $\langle
A\rangle_{m}=\epsilon(\phi)\; \nu/\lambda$ [see
Eqs.(\ref{eq:maxpos},\ref{eq:maxval})]. In the range $\gamma=[0,3]$,
the dotted and the full line practically coincide. We assumed
$\lambda/\delta p = 0.01,  \lambda/p_\phi = 0, \delta p=\Delta P
=p_H$.}
\end{figure}

The exact average value is
\begin{align}
\nonumber \langle A\rangle =&
\biggl\{ \mathbf{n}\!\cdot\!\mathbf{n}_{i}
+\mathbf{n}\!\cdot\!\mathbf{n}_{f} -\kappa^2 e^{-\lambda^2/2\nu^2}
\biggl[\cos{\!\left(\!\frac{\lambda}{p_\phi}\!\right)}
\mathbf{n}\!\cdot\!(\mathbf{n}_{i}\!\times\!\mathbf{n}_{f})
 \\
& \label{eq:exactval} -\sin{\left(\frac{\lambda}{p_\phi}\right)}
(\mathbf{n}\!\times\!\mathbf{n}_{i})\cdot
(\mathbf{n}\!\times\!\mathbf{n}_{f}) \biggr] \biggr\}\big/N.
\end{align}

To lowest order in $\lambda$, $\langle A\rangle$ is given by
Eq.~\eqref{eq:genweakval2} with $A_{w} = \mathbf{n}\cdot \left[
\mathbf{n}_{i} +\mathbf{n}_{f} +i \mathbf{n}_{i}\!\times\!\mathbf{n}_{f}
\right]/\left[1+\mathbf{n}_{i}\cdot\mathbf{n}_{f}\right]$.
Interestingly, when $\mathbf{n}$ lies in the plane orthogonal to the
bisector of $\mathbf{n}_{i}$ and $\mathbf{n}_{f}, $ $A_{w}$ is purely
imaginary.
This setting of the weak measurement can hence be a testing ground
to detect the contribution of the imaginary part.

Without loss of generality, we take $\mathbf{n}_{i}, \mathbf{n}$ to
define the $XZ$ plane, with the former as the $Z$-axis, and the
latter forming an angle $\theta\in[0,\pi]$.
The direction $\mathbf{n}_{f}$ is defined by the azimuthal and polar
angles $\gamma \in[0,2\pi], \phi\in[0,\pi]$ \footnote{Notice that we
are at variance with the standard convention $\gamma \in[0,\pi],
\phi\in[0,2\pi]$.}. Then $A_{w}=\cos{\theta}+\sin{\theta} e^{-i\phi}
\tan{(\gamma/2)}$. In Figure \ref{fig:example} we plotted $\langle
A\rangle$ as a function of $\gamma$ for fixed $\theta,\phi$. We
compare the exact value
[Eqs.~(\ref{eq:exactval},\ref{eq:normalizing})],
the approximate value $\langle A \rangle_0$ given by
Eq.~\eqref{eq:genweakval2}, and ${Re}(A_{w})$.
For fixed $\theta\neq 0,\pi$, and $\phi$, $\langle A\rangle$ reaches
extremal values for $\gamma^*=\pi-\eta^*$,
\begin{equation}\label{eq:maxpos}
\eta^* \simeq \lambda
\frac{\sin{\theta}}{\varepsilon(\phi)}\left[\frac{\textstyle
\kappa^2}{\textstyle p_\phi} \pm \sqrt{ \frac{\textstyle
\varepsilon(\phi)^2}{\textstyle \nu^2}+\frac{\textstyle (1+\kappa^4)
\left(\cos{\phi}\right)^2}{\textstyle p_\phi^2}
}\right] ,
\end{equation}
with $\varepsilon(\phi):=\cos{\phi}+\kappa^2\sin{\phi}$.
The extremal value is
\begin{equation}\label{eq:maxval}
\langle A\rangle_{m} = \frac{
\frac{\textstyle\varepsilon(\phi)^2}{\textstyle \lambda}} {
\frac{\textstyle \cos{\phi}\; \varepsilon'(\phi) }{\textstyle
p_\phi} \pm \sqrt{ \frac{\textstyle \varepsilon(\phi)^2}{\textstyle
\nu^2}\!+\!\frac{\textstyle (1+\kappa^4)
\left(\cos{\phi}\right)^2}{\textstyle p_\phi^2}
} },
\end{equation}
with the prime meaning differentiation. 
\begin{figure}[t!]
\includegraphics[width=2.3in]{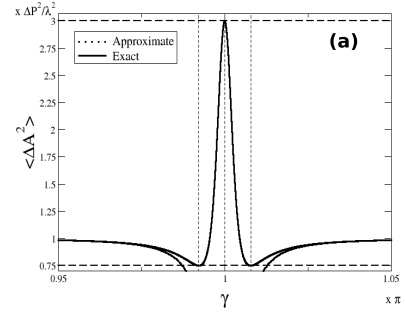}\\[-25mm]
\includegraphics[width=2in]{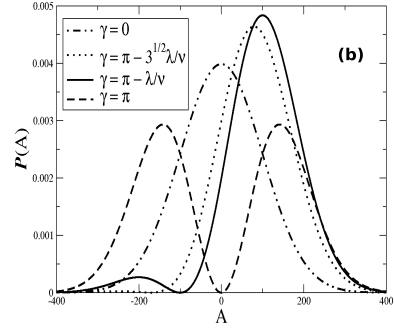}\\[-25mm]
\caption{\label{fig:spread} (a) The variance of $A$
as a function of $\gamma$ for fixed $\theta=\pi/2$, $\phi=\pi/4$.
The dashed vertical and horizontal lines correspond to the positions
and the values of the extrema [Eqs.
(\ref{eq:spreadmin},\ref{eq:spreadmax})]. (b) The
probability distribution for $\theta=\pi/2$, $\phi=\pi/4$ for some
significant values of $\gamma$. The parameters are the same as those
of Fig.~\eqref{fig:example}. }
\end{figure}
Generally, the upper sign solution in
Eqs.~(\ref{eq:maxpos},\ref{eq:maxval}) holds as far as
$\varepsilon(\phi)\gg \lambda/p_\phi,\lambda/\nu$, in which case the extremum is no
longer found close to $\pi$, but
\begin{equation}
\sin{\gamma^*} \simeq \frac{ -2\kappa^2\frac{\lambda}{p_\phi}
\sin{\theta}
\left(\varepsilon(\phi)+\varepsilon'(\phi)\frac{\lambda}{p_\phi}\cos{\theta}\right)}
{\left(\kappa^2\frac{\lambda}{p_\phi} \sin{\theta}\right)^2+
\left(\varepsilon(\phi)+\varepsilon'(\phi)\frac{\lambda}{p_\phi}\cos{\theta}\right)^2
} \;,
\end{equation}
and $\langle A\rangle_m\simeq \cos{\theta}$, while the lower sign
solution converges to $\eta^* \simeq -\lambda
\sin{\theta}/p_\phi\sqrt{1+\kappa^4}$,
\[\langle A\rangle_{m} = -\frac{2\kappa^2\lambda/p_\phi}{\kappa^4\lambda^2/(1+\kappa^4)p_\phi^2+\lambda^2/\nu^2}
.\]
There are two exceptions to this: (\emph{i}) For $p_\phi\gg \nu$, $\eta^*
\simeq \pm\sin{\theta}{\lambda}/{\nu}$, and $\langle A\rangle_{m} =
\pm\varepsilon(\phi){\nu}/{\lambda}$. (\emph{ii}) For $\kappa^2=
2\Delta P^2 \Delta t/\hbar M_{p} \ll 1$, $\eta^* \simeq
\pm\sin{\theta} \lambda\sqrt{1/\nu^2+1/p_\phi^2}$, and $\langle
A\rangle_{m}\!=\!
 \cos(\phi)/ \lambda\left[\pm\sqrt{1/\nu^2+1/p_\phi^2}-\sin{\phi}/p_\phi\right]$.

The extremal value for $\langle A \rangle$ as a function of both
$\gamma,\phi$ has an involved expression, except for $p_\phi\gg
\nu$, when the location of the extremum is
$\gamma^*=\pi\mp\sin{\theta}\lambda/\nu, \phi^* =
{Arctan}(\kappa^2)$, and $\langle A
\rangle_{m}=\pm\sqrt{1+\kappa^4} \nu/\lambda$. In the same limit, we
have that the minimum of the spread is reached for $\eta^*\simeq
\pm\sqrt{3}\sin{\theta}\lambda/\nu, \phi^* =
{Arctan}(\kappa^2)$
\begin{equation}\label{eq:spreadmin}
\langle \Delta A^2\rangle_{min} = \frac{\Delta P^2-(1+\kappa^4)
\nu^2/4}{\lambda^2}\ge \frac{3}{4}\frac{\Delta P^2}{\lambda^2} .
\end{equation}
Its maximum is reached for $\gamma^*=\pi$, and it is
\begin{equation}\label{eq:spreadmax}
\langle\Delta A^2\rangle_{max} \simeq \frac{\Delta P^2+2(1+\kappa^4)
\nu^2}{\lambda^2}\le 3\frac{\Delta P^2}{\lambda^2} .
\end{equation}
We plot the probability distribution for three values of $\gamma$:
close to $\gamma=\pi$ the distribution has two peaks, each of order
$100$ for the choice of parameters made. While the average value
goes to zero when $\gamma$ gets very close to $\pi$ ($\pi-\gamma\ll
\lambda/\nu$), the probability density of observing a value in the
range $[-1,1]$ is rather small: in each individual measurement, it
is likely that the value of $|A|$ will be much larger than unity.

\section{Conclusions}
We have showed that accounting for the dynamics of the probe in the
weak measurement leads to an observable deviation of the average
value from the real part of the complex weak value defined in
Ref. \cite{AAV}. We have also derived an expression for the spread,
and, in the case of spin, we have individuated the locations and
values of the extrema of $\langle A \rangle$.

This work was supported by FAPESP and CNPq.

\end{document}